\documentclass[]{aa}
\usepackage{graphicx}
\usepackage{txfonts}
\begin{document}
  \title{The LAEX and NASA Portals for CoRoT Public Data \thanks{CoRoT is a space project operated by the French Space Agency,
CNES, with participation of the Science Programme of ESA,
ESTEC/RSSD, Austria, Belgium, Brazil, Germany, and Spain.}}


  \author{E. Solano\inst{1,11} \thanks{e-mail: esm@laeff.inta.es}
         \and
         K. von Braun\inst{2,9}
        \and
         A. Velasco\inst{1,11}
        \and
         D. R. Ciardi\inst{2,9}
        \and
	  R. Guti\'{e}rrez\inst{1,11}
	  \and
        D. L. McElroy\inst{2,9}
	  \and
	  M. L\'opez \inst{1,11}
         \and
        M. Abajian\inst{2,9}
       \and
        M. Garc\'{\i}a\inst{3,11}
	  \and
        B. Ali\inst{4,9}
         \and
	  L. M. Sarro\inst{5,11}
         \and
        G. B. Berriman\inst{2,9}
       \and
	 G. Bryden\inst{6}
       \and B. Chan\inst{2,10}
       \and J. Good\inst{2,9}
       \and S. R. Kane\inst{2,9}
       \and A. C. Laity\inst{2,9}
      \and C. Lau\inst{2,10}
       \and A. N. Payne\inst{7}
       \and P. Plavchan\inst{2,9}
      \and S. Ramirez\inst{2,9}
       \and M. Schmitz\inst{4,9}
       \and J. R. Stauffer\inst{8,9}
       \and P. L. Wyatt\inst{2,9}
       \and A. Zhang\inst{2,4,9}
         }

  \institute{Centro de Astrobiolog\'{\i}a, CSIC-INTA. LAEX. P.O. Box 78. 28691 Villanueva de la Ca\~{n}ada, Madrid, Spain
        \and
            NASA Exoplanet Science Institute
	 \and
	     \'{A}rea de Lenguajes y Sistemas Inform\'{a}ticos. Universidad Pablo Olavide. Ctra. de Utrera, km. 1, 41013 Sevilla, Spain
	 \and
	     Infrared Processing and Analysis Center
	\and
	     Dpt. de Inteligencia Artificial, UNED, Juan del Rosal, 16, 28040 Madrid, Spain
	\and
	     Jet Propulsion Laboratory, 4800 Oak Grove Drive, Pasadena, CA 91109, USA
	\and
	     Australian National University, Mount Stromlo Observatory, Canberra, ACT, Australia
\and
            Spitzer Science Center
\and
            California Institute of Technology, 770 South Wilson Ave, Pasadena, CA 91125, USA
\and
	     Raytheon Information Systems, Pasadena, CA, USA
\and
	     Spanish Virtual Observatory Thematic Network}

  \date{Received ; accepted}


 \abstract
  {}
{We describe here the main functionalities of the LAEX (Laboratorio de Astrof\'{\i}sica Estelar y Exoplanetas /Laboratory for Stellar Astrophysics and Exoplanets) and NASA portals for CoRoT Public Data. The CoRoT
archive at LAEX was opened to the community in January 2009 and is managed in the framework of the Spanish Virtual Observatory. NStED (NASA Star and Exoplanet Database) serves as the CoRoT portal for the U.S. astronomical community. NStED is a general purpose stellar and
exoplanet archive with the aim of providing support for NASA's planet finding and characterisation goals, and the planning and support of NASA and other space missions. CoRoT data at LAEX and NStED can be accessed at http://sdc.laeff.inta.es/corotfa/ and http://nsted.ipac.caltech.edu, respectively.}
{Based on considerable experience with astronomical archives, the aforementioned archives are designed with the aim of delivering science-quality data in a simple and efficient way.}
{LAEX and NStED not only provide access to CoRoT Public Data but furthermore serve a variety of observed and calculated astrophysical data. In particular, NStED provides scientifically validated information on stellar and planetary data related to the search for and characterization of extrasolar planets, and LAEX makes any information from Virtual Observatory services available to the astronomical community.}
  {}


  \maketitle
%

\section{Introduction}

The ultimate goal of any scientific mission is to obtain data producing new discoveries and new science. As astronomical research is expanded through the design and execution of innovative and groundbreaking missions, there is increasing awareness by scientists and funding agencies of the need to maximize scientific return on these significant financial investments. Astronomical archives play a fundamental role in ensuring this return and are essential tools for modern astronomical research as demonstrated by the intensive usage by the scientific community that they enjoy. This paper describes two of the CoRoT Public Archives: LAEX and NStED.

LAEX (formerly known as LAEFF and belonging to Centro de Astrobiolog\'{\i}a) has a long history in astronomical archives. In 1998, it was selected to host the INES Final Archive at the end of the International Ultraviolet Explorer (IUE) project. Much of the experience acquired with INES was used to develop GAUDI (\cite{sol05}), the starting point of LAEX's participation in CoRoT. The GAUDI system was designed to conduct ground-based observations obtained in preparation of the CoRoT asteroseismology program and make them available in a simple and efficient way.

In carrying out scientific research, collecting ancillary information about
science targets requires thorough reviews of the literature across many
different wavelengths and observational techniques.  The NASA Star and
Exoplanet Database (NStED) was created to address the lack of a central
repository of such multi-dimensional observational data and to provide to the
community a straightforward method to gather scientifically-validated
information on nearby and exoplanet-hosting stars for their research.  NStED
is principally dedicated to collecting and serving published data involved in
the search for and study of extrasolar planets and their host stars.

CoRoT\footnote{http://corot.oamp.fr} (Convection, Rotation and Planetary Transits) is a mission with two principal objectives: the study of stellar interiors using asteroseismology techniques, and the discovery of extrasolar planets using the transit method. Successfully launched in December 2006, CoRoT has been providing the astronomical community with a number of revolutionary results on the internal structure of stars and on the physical properties of extrasolar planets.

The need for an archive for the CoRoT data was identified in the early phases of the project. In March 2003, LAEX was selected, together with the Centre de Donn\'ees Astronomiques de Strasbourg (CDS)\footnote{http://cdsweb.u-strasbg.fr/}, to be responsible for the long-term storage and maintenance of the CoRoT Final Archive, which will contain
all CoRoT data processed in a homogeneous and uniform way. This Final Archive will be Virtual Observatory compliant. Furthermore, it is designed to be a static archive (its contents will not change with time) that will represent the legacy of the mission for future generations of researchers and educators.

In addition to the Final Archive, efficient data management and data interchange mechanisms have been set up to work during the operational phase of CoRoT. In this framework two types of archives have been defined.

\begin{itemize} 
\item{\it The Mission Archive}: Located at the Institut d'Astrophysique Spatiale (IAS) in Paris, the Mission Archive is the only access point for CoRoT data during the proprietary period.\\
\item{\it The Public Archives}: Containing only publicly available data, these are dynamic archives since new data as well as re-calibrations of previous data will be ingested during the mission lifetime. The Public Archives are being developed at the IAS (Paris), CDS (Strasbourg), NStED (Pasadena, California), and LAEX (Madrid). For the sake of integrity, the data are stored at IAS and remotely accessed from the Public Archives.
\end{itemize}

This paper gives descriptions the two Public Archives NStED (\S \ref{nsted}) and LAEX (\S \ref{laex}) and summarizes in \S \ref{summary}.

\section{NStED}\label{nsted}

NStED\footnote {http://nsted.ipac.caltech.edu} is divided into two principal components: the stellar and exoplanet services (\S \ref{stellar}) and
the exoplanet transit survey service (ETSS; \S \ref{etss}).

The stellar and exoplanet services provide access to stellar parameters of
bright stars and exoplanet hosting stars, along with exoplanet parameters
wherever available. The stellar service includes the following:
\begin{itemize}
\item Stellar parameters for $\sim$140,000 bright nearby stars.
\item Enable queries for individual stars, or search by stellar/planetary
parameters.
\item Associated published images, spectra, and time series data.
\end{itemize}
Complementary to the stellar service is the exoplanet service, which includes
the following:
\begin{itemize}
\item General data and published parameters for the known exoplanets and host stars,
updated weekly.
\item Photometric and radial velocity data related to the known exoplanets,
including multi-instrument time-series.
\end{itemize}

The purpose of NStED-ETSS is to make available to the astronomical community
full sets of time-series data (i.e., light curves) of planet transit studies
and other variability surveys in a homogeneous format, along with tools for
data analysis and manipulation. Principal goals of NStED-ETSS include the
following:
\begin{itemize}
\item Provide access to support data for ground-based and space-based missions.
\item Allow the development of different or improved algorithms for transit
detection or variability classification on complete existing survey data sets;
for instance, to enable the detection of planets previously missed in the
original study.
\item Extend the time baseline for transit studies by using data sets containing the
same stars, leading to increased detection efficiency, results of increased
statistical significance, enhanced potential to conduct transit timing
studies, etc.
\item Enable improved understanding of false positivies encountered in transit surveys.
\item Provide access to a wealth of other astrophysical results and ancillary
science not pursued in the original survey, such as studies of eclipsing
binary and other variable stars or variability phenomena, stellar atmospheres
(rotation, flares, spots, etc.), asteroseismology and intrinsic stellar
variability, as well as serendipitous discoveries such as photometric
behaviours of supernovae progenitors, etc.
\end{itemize}

These two NStED services are described in Sect. 2.1 and 2.2, and a discussion of how NStED will act as the U.S. portal to the CoRoT data is given in Sect. 2.3.
\subsection{{The NStED Stellar and Exoplanet Services}}\label{stellar}

\begin{figure}[h]
   \includegraphics[width=\columnwidth]{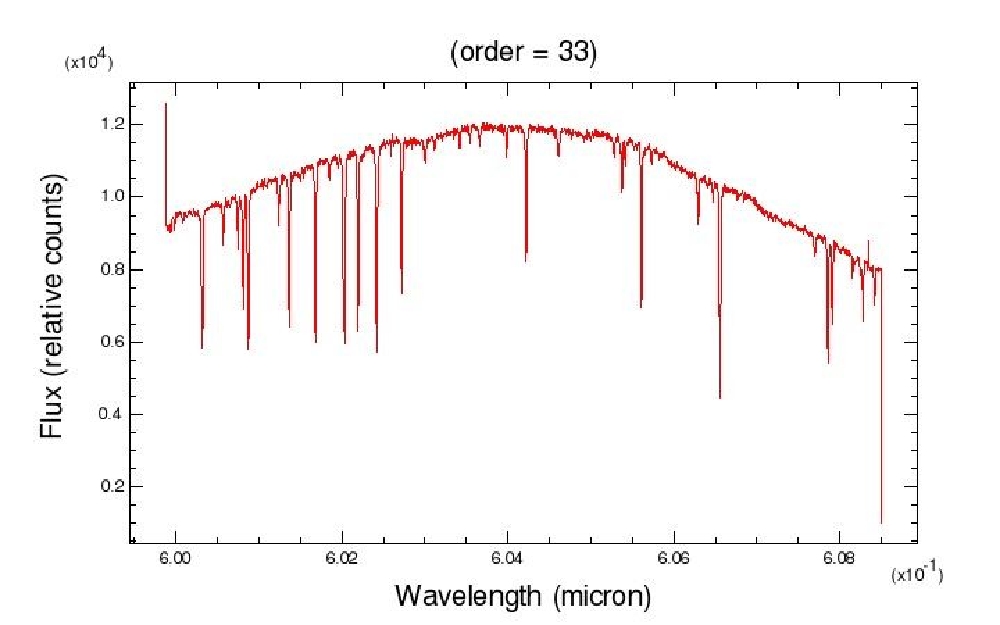}
   \caption{Example of data in NStED's stellar services: N2K spectrum of HD
   804 (\cite{fis05}).}  \label{fig_spectrum}
\end{figure}

The NStED stellar and exoplanet content (\cite{rab09}) is composed of
published tabular data, derived and calculated quantities, and associated data
including images, spectra, and time series.  NStED's core set of $\sim$140,000
stars (out of a total of approximately 500,000 stars) is derived from the Hipparcos (\cite{p97}), Gliese-Jahreiss (\cite{gj91}), and Washington Double Star (\cite{mwh01}) catalogs. A summary
of the stellar parameters and associated data in NStED is shown in Table
\ref{tab_stellar}. An example spectrum from the N2K consortium (\cite{fis05})
contained in NStED is shown in Figure \ref{fig_spectrum}.

NStED currently supports complex multi-faceted queries on astrophysical stellar and exoplanet parameters (Tables \ref{tab_stellar} and
\ref{tab_exoplanets}).
Queries to NStED can be made using constraints on any combination of these
parameters. In addition, NStED provides tools to derive specific inferred
quantities for the stars in the database, cross-referenced with available
extrasolar planetary data for those host stars.

\begin{figure}[h]
   \includegraphics[width=\columnwidth]{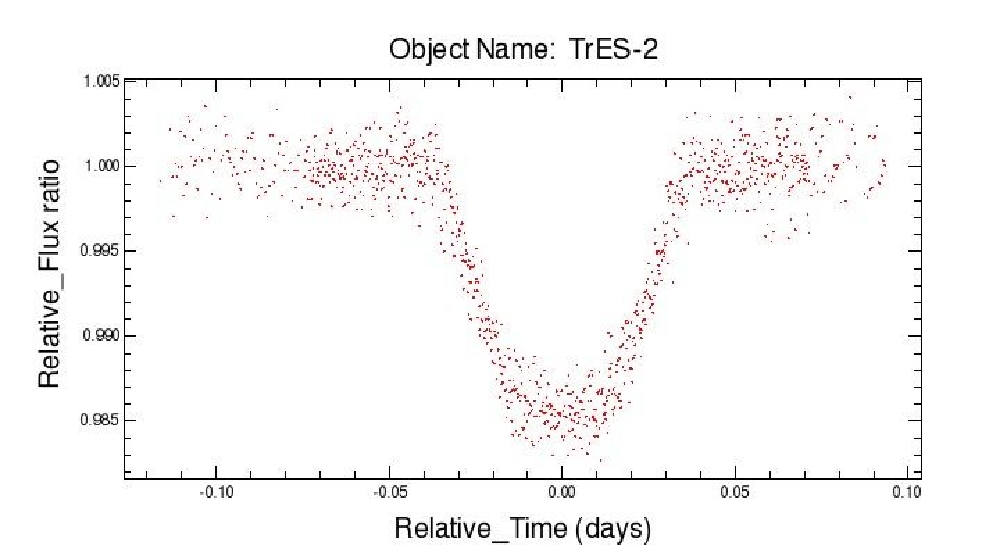}
   \caption{Example of time-series data contained in NStED's exoplanet services:
   light curve of the transiting exoplanet TrES-2 (\cite{odo06,hwl07}).}
   \label{fig_tres2}
\end{figure}

In order to facilitate future exoplanet studies, NStED maintains an up-to-date
list of all known exoplanetary systems and associated stellar data by daily
monitoring the astronomical literature and making weekly updates to the
database. Available data in NStED include high-precision light curves (e. g.,
the light curve of the transiting planet TrES-2 shown in Figure
\ref{fig_tres2}). Predicted observable signatures of exoplanets are calculated
to aid users in selection of stars appropriate for planet searching and
characterisation. The exoplanet signature predictions include habitable zone
sizes, astrometric and radial velocity wobbles, and transit depths.  A summary
of the exoplanet parameters and associated data within NStED is shown in Table
\ref{tab_exoplanets}. Fig. \ref{fig_aitoff} features a graphic display of the
sky distribution of stars and data in NStED.

In the near future, NStED will furthermore feature planet transit ephemerides
and tools to predict transit observability windows for given telescope
locations, courtesy of G. Laughlin's site {\tt http://transitsearch.org}.

\begin{table}
 \begin{center}
   \caption{Summary of stellar parameters available in NStED.}
 \label{tab_stellar}
   \begin{tabular}{@{}lll}
     \hline
     Published Parameters & Derived Parameters & Associated Data \\
     \hline
     Position, Distances & Temperature      & 2MASS Images  \\
     Kinematics          & Luminosity       & N2K Stellar Spectra \\
     Photometry, Colors  & Radius           & Coronographic Images\\
     Spectral Type       & Mass             & MOST Light Curves\\
     Luminosity Class    & LSR Space Motion & Hipparcos Light Curves\\
     Metallicity         & & Ground-based Light Curves\\
     Rotation            & & \\
     Activity Indicators & & \\
     Variability         & & \\
     Multiplicity        & & \\
     \hline
   \end{tabular}
 \end{center}
\end{table}

\begin{table}
 \begin{center}
   \caption{Summary of exoplanetary parameters available in NStED.}
       \label{tab_exoplanets}
   \begin{tabular}{@{}lll}
     \hline
     Published Parameters & Predicted Parameters & Associated Data \\
     \hline
     Number of Planets & Habitable Zone          & RV Curves \\
     Planetary Mass    & Astrometric Signature      & Light Curves  \\
     Orbital Parameters    & Radial Velocity Signature  & High-Contrast\\
     Link to entry in & Transit Depths & Images \\
     {\tt http://exoplanet.eu} &  & \\
     \hline
   \end{tabular}
 \end{center}
\end{table}

\subsection{The NStED Exoplanet Transit Survey Service}\label{etss}

\begin{figure*}[h]
\begin{center}
\includegraphics[width=12cm,clip=true]{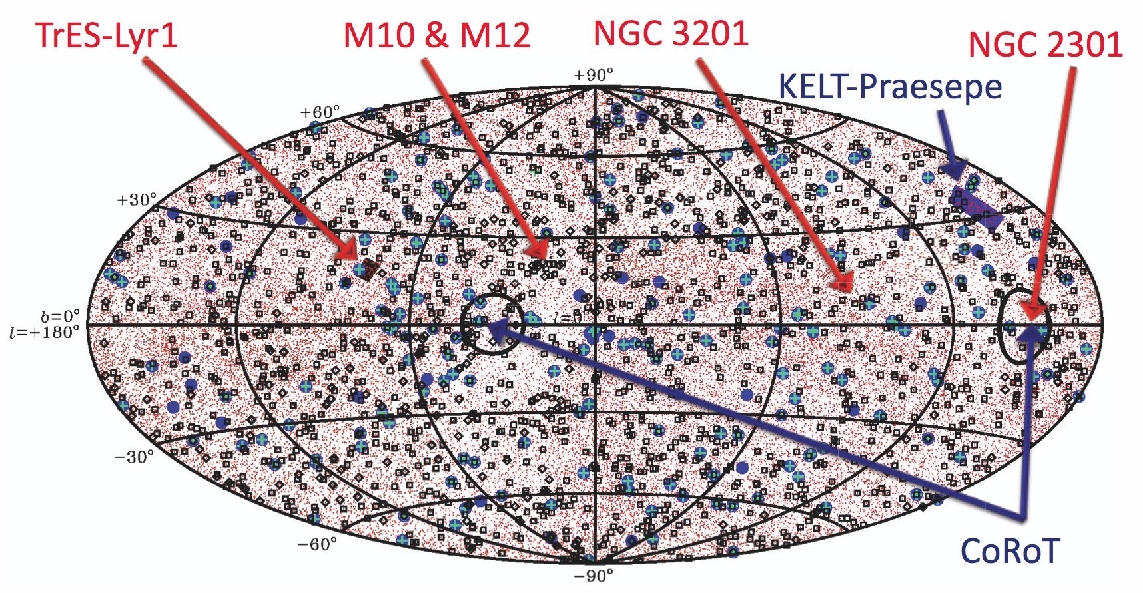}
\caption{NStED Contents: Aitoff projection with contents of NStED's
stellar service and the locations of the ETSS survey data sets.  Small dots:
dwarf stars (for clarity, the giant stars are not plotted); large dots:
exoplanet hosting stars; large plus signs: stars with radial velocity curves
or photometric lightcurves; open squares/diamonds: stars with
images/spectra. The locations of the ETSS data sets are shown by arrow,
including the two CoRoT ``eyes''.}  \label{fig_aitoff}
\end{center}
\end{figure*}

NStED's exoplanet and transit survey service (\cite{baa09}) currently contains
time-series data (light curves) of all monitored stars in four variability
surveys of the stellar clusters M10, M12, NGC 2301, and NGC 3201
(\cite{hvt05,the05,bm02,bmc02}), and two wide-field exoplanet transit surveys
(TrES-Lyr1: \cite{odo06}, and KELT-Praesepe:
\cite{ppd07,psp08}). Fig. \ref{fig_aitoff} shows the locations of these data
sets in an Aitoff projection of the celestial sphere.

Each data set contained in ETSS features a master file and a single file for
each light curve. Tools enable the user to visualize the data and perform
manipulation and analysis tasks.

The master file provides basic properties of the data set as a whole as well
as global parameters about each individual light curve file. Through the NStED
infrastructure, one may query the master file to search the data set by
parameters such as unique identifiers, celestial coordinates, static
photometry parameters (single-epoch magnitudes), variability filter(s),
Heliocentric Julian Dates (HJD), number of observational epochs, root-mean-square
dispersion about the median magnitude, median absolute deviation, existence
(and frequency) of photometric outliers, $\chi^2$ about the median magnitude,
cross identification between different magnitudes, etc.

Each light curve file is associated to a unique identifier and features a
header summarizing global information about the light curve, as well as the
column-delimited photometry data, generally in the format HJD, magnitude,
uncertainty. Thus, it is flexible and readable with all computer operating
systems and can be translated to other formats such as Virtual Observatory (VO), binary FITS tables,
etc.

\begin{figure}[h]
\includegraphics[width=\columnwidth]{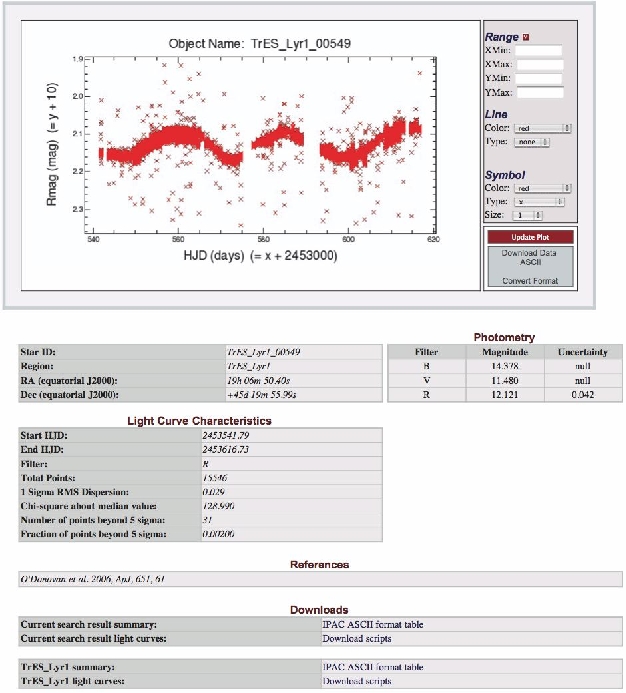}
\caption{ETSS Detail Page in NStED: featured are an interactive light curve viewer
(mag vs HJD), summary of light curve characteristics, direct links to ascii
light curve, cross-identified stars (if applicable), summary table, download
scripts, and the data set summary (master file).}  \label{fig_example}
\end{figure}

Fig. \ref{fig_example} shows an example of data visualization found on the
NStED-ETSS website, complete with light curve characteristics, data set
reference, and links to the associated files and download scripts. The data in
this plot are taken from the TrES-Lyr1 data set (\cite{odo06}).  Finally, data
sets to be ingested in the near future include XO-1 (\cite{msv06}) and WASP0
(\cite{kch04}).

\subsection{NStED and CoRoT}\label{corot}

Through its existing infrastructure, NStED is optimally suited to serve as
portal to the public CoRoT data (Fig. \ref{fig_aitoff}) for the astronomical
community in the United States. The corresponding archive for the CoRoT data,
a collaboration between NStED and CNES/ESA, is being developed in concert with
the CoRoT mission archives at IAS and the archives at LAEX and CDS. Principal
goals of this collaboration include establishing a comprehensive archive to
serve the CoRoT light curves, provide interconnectivity between CoRoT and
NStED's ground-based data, and develop visualization and characterization
tools for light curve analysis and manipulation.

\begin{figure}[h]
\includegraphics[width=\columnwidth]{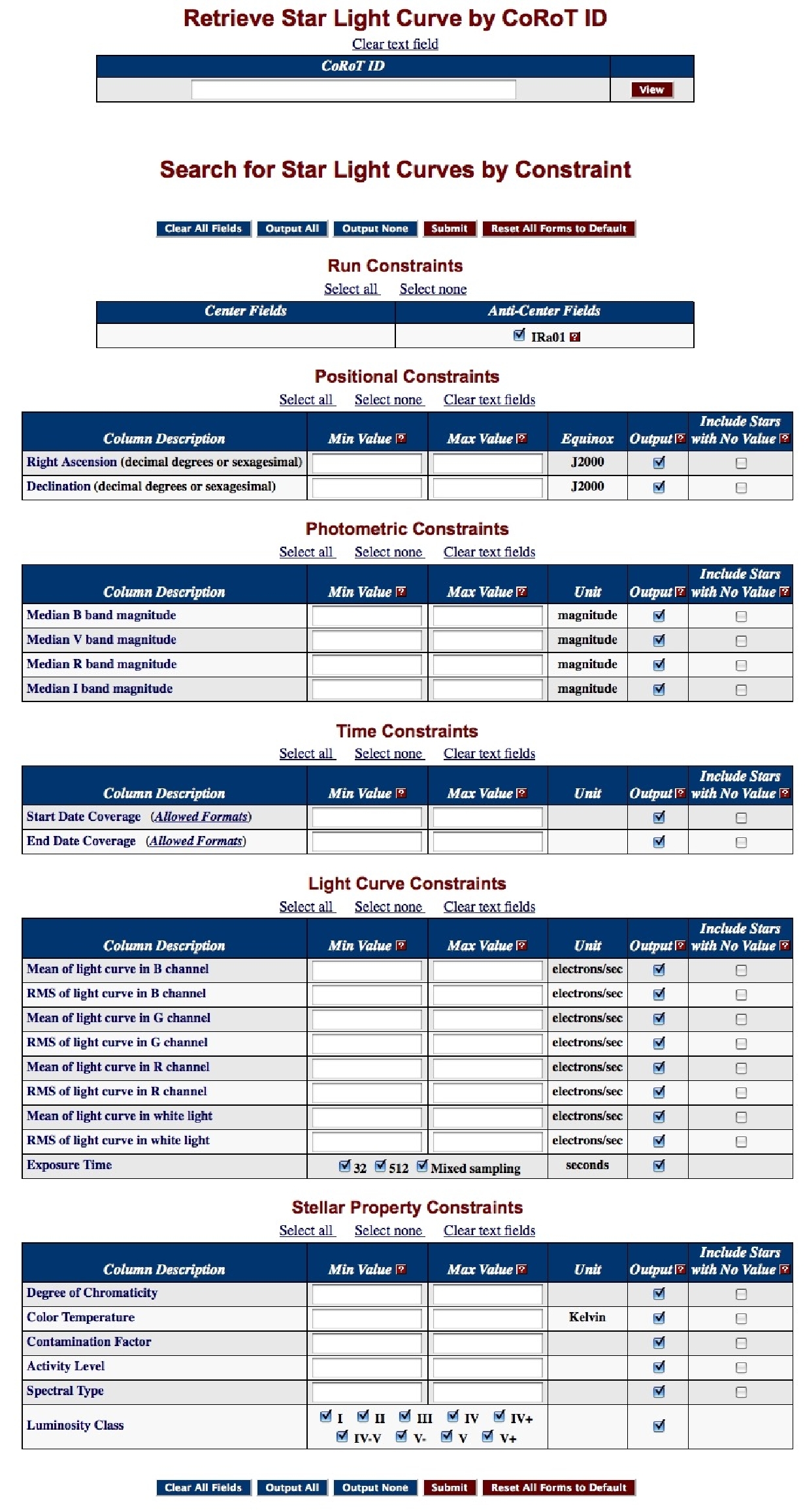}
\caption{NStED page to query CoRoT exoplanet data. The user may search a
specific CoRoT star (by ID number), search by positional or photometry
criteria, by observational epochs, by time-series or stellar properties, and
across multiple CoRoT runs. Returned are all data that fulfill the
combination of the entered criteria. The user may then obtain all thus
selected data or look at individual stars, an example of which is shown in
Fig. \ref{fig_visualizer}. The equivalent page exists for the
asteroseismology field. } \label{fig_nsted_exo}
\end{figure}

\begin{figure}[h]
\includegraphics[width=\columnwidth]{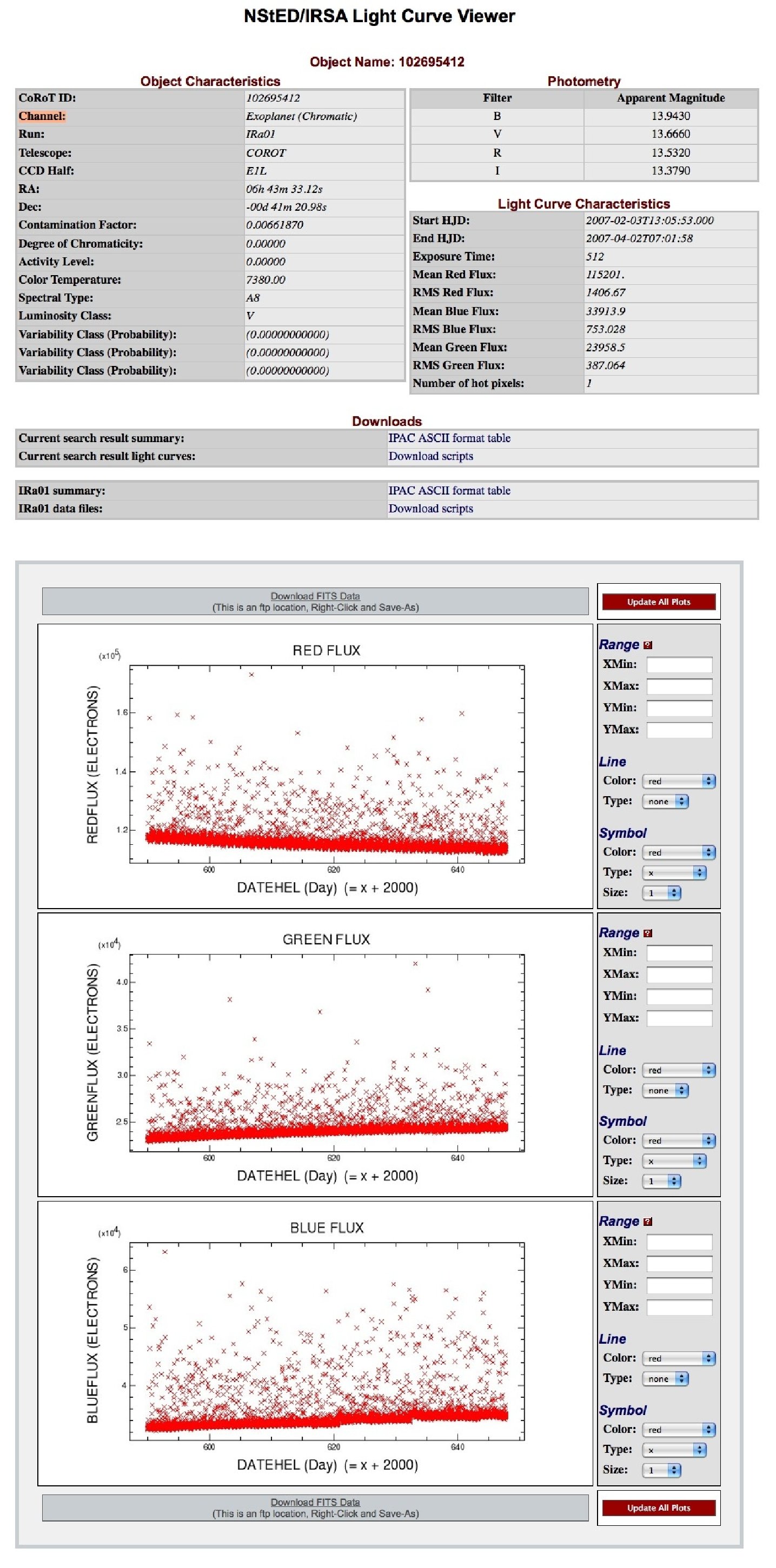}
\caption{Example of NStED visualizer page of an individual CoRoT star in the
exoplanet field. Header information is given at the top, along with links to
download options. The customizable visualizer shows the CoRoT time-series
data in the three individual CoRoT colors. A similar page is available for
asteroseismology targets.}  \label{fig_visualizer}
\end{figure}

The main aspects of the NStED interface to the CoRoT data are the following
(see Figures \ref{fig_nsted_exo} and \ref{fig_visualizer}):
\begin{itemize}
\item Separate and independent interfaces for the exoplanetary and
asteroseismology fields.
\item Ability to search by astrophysical or observational parameters, or
alternatively by CoRoT ID number.
\item Ability to search across multiple CoRoT runs.
\item Direct links to individual CoRoT seismology targets and option of
accessing any available NStED data for the target.
\item Tabular results can be saved for offline analysis.
\item Visualization page for individual CoRoT
targets (Fig. \ref{fig_visualizer}).
\item Download scripts to obtain either all light curves in a given CoRoT run,
or to obtain only the light curves that fulfill the specified criteria
(Fig. \ref{fig_nsted_exo}).
\end{itemize}

\section{The CoRoT Public Archive at LAEX} \label{laex}

\subsection{Functionalities: Archive search}

\begin{figure*}[h]
  \centering
\includegraphics[width=12cm,clip=true]{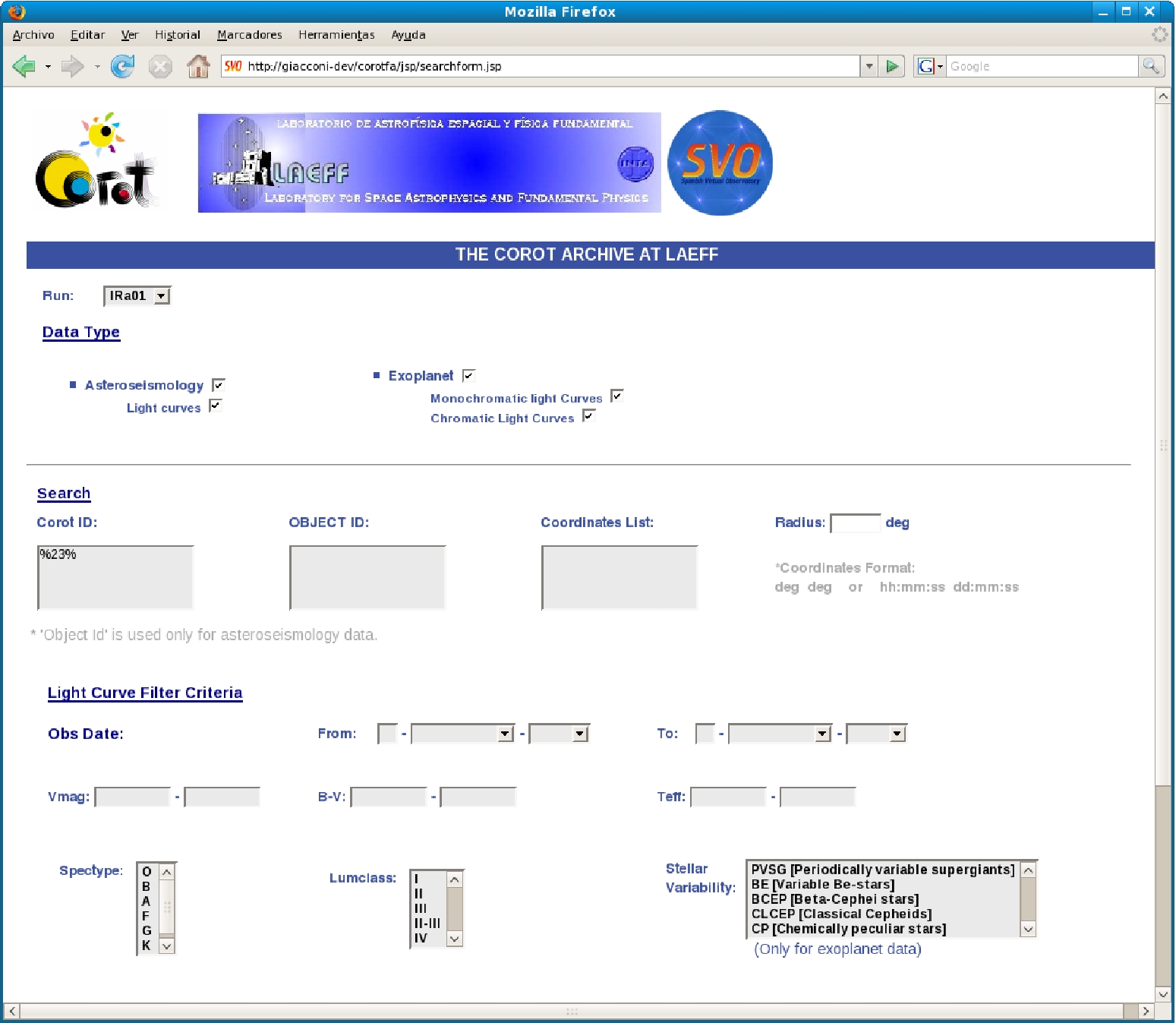}
  \caption{The CoRoT Public Archive at LAEX. Input query form.}
             \label{input}%
   \end{figure*}

The CoRoT archive at LAEX\footnote{http://sdc.laeff.inta.es/corotfa/} is accessed by means of a web-based fill-in form that permits
queries by observing run, observational programme (asteroseismology
or exoplanets), type of data (monochromatic or chromatic light curves in the exoplanet channel), CoRoT identification, object name,
and coordinates and radius. Searches can be filtered by different criteria like
observing date, $V$ magnitude, ($B-V$) color, effective temperature, spectral type and/or luminosity
class. Searches are case-independent, and wildcards (\%) are
permitted for the CoRoT ID field. The system furthermore incorporates a built-in name resolver allowing queries
by any of the names provided by the SIMBAD\footnote{http://simbad.u-strasbg.fr/simbad/} database. The output fields may be
ordered by coordinates or CoRoT identifier. The output format is in HTML with a pre-defined number of results shown per page (Fig.~\ref{input}).

\subsection{Functionalities: Results from search}
\begin{figure*}[t]
  \centering
\includegraphics[width=12cm,clip=true]{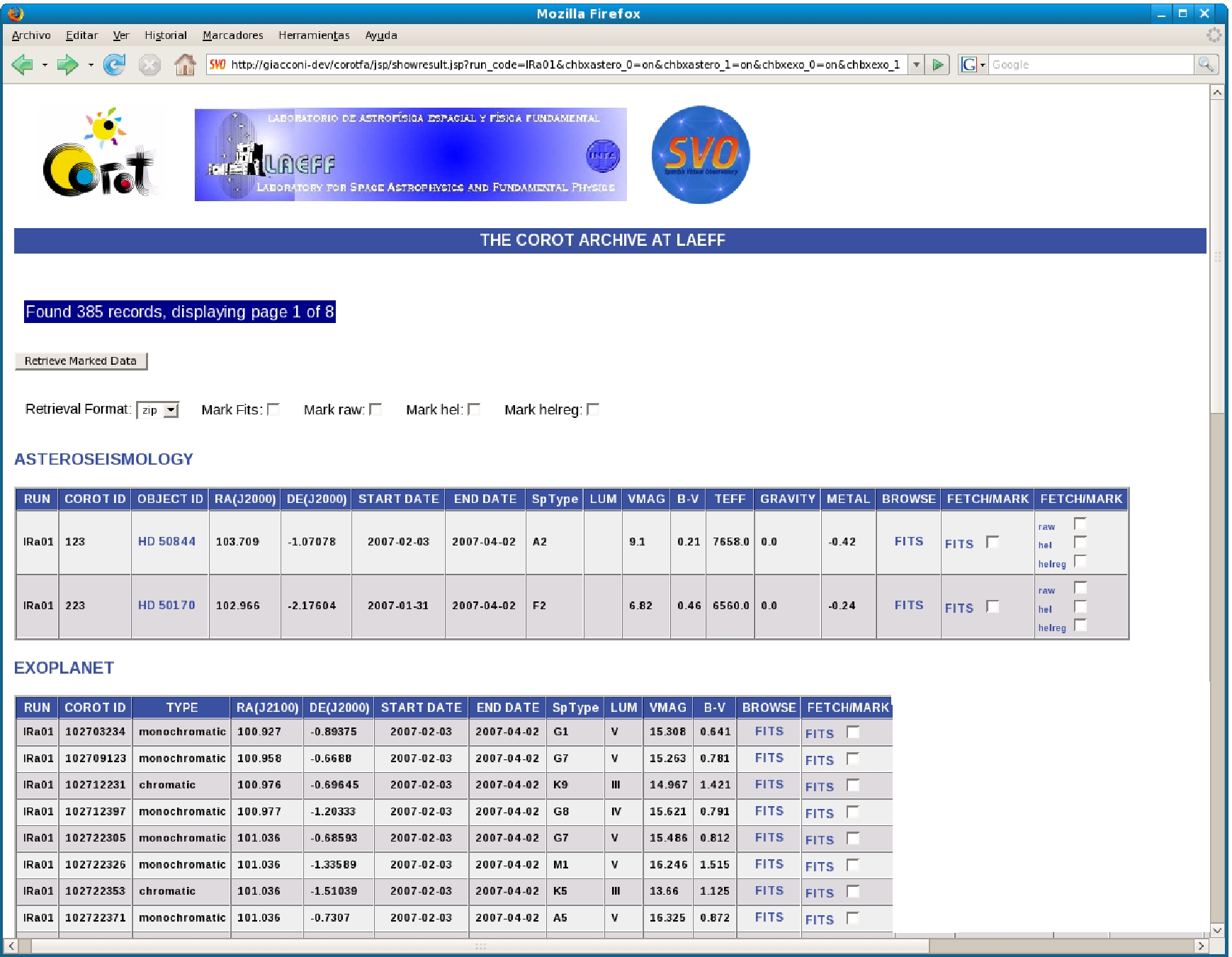}
  \caption{The CoRoT Public Archive at LAEX. Output of a query.}
             \label{output}%
   \end{figure*}

\begin{figure*}[h]
  \centering
\includegraphics[width=12cm,clip=true]{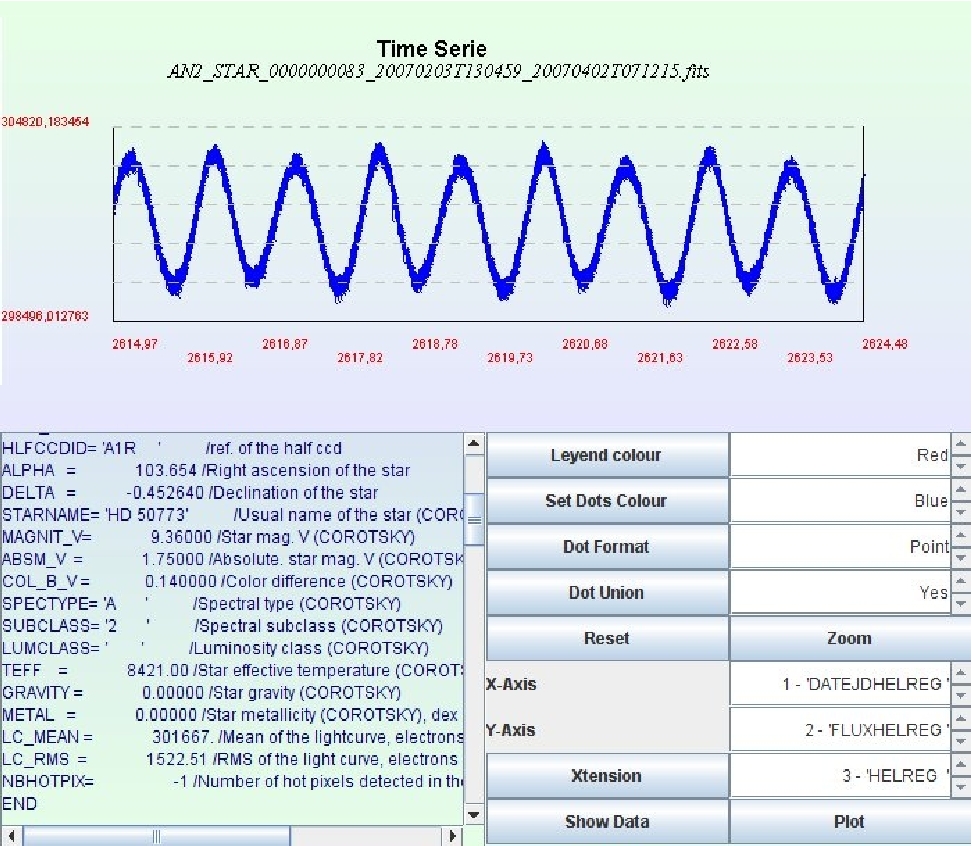}
\caption{Light curve and FITS header visualisation.}
\label{plot}
\end{figure*}

An example of the result of a query (as described above) is given in Fig.~\ref{output}. Data of interest can be retrieved
as FITS files. In addition, the following options are provided:

\begin{itemize}
\item Name resolver: By clicking the object name, a list of equivalent names as
provided by SIMBAD is displayed. \\

\item Multidownload: Data in FITS format can be retrieved in groups. For multiple-file retrieval
it is possible to include or exclude individual datasets. Multiple-file download
generates a file in ZIP format. For asteroseismology data, the system also offers the possibility of downloading each of the FITS binary table extensions (Raw, Hel and Helreg) in ASCII format.\\

\item FITS Header/Data preview: An interactive plot of the light curve as well as a visualization
of the associated FITS header is generated by clicking on the corresponding
link. The user can select the columns and the FITS extension (for the asteroseismology programme) to be plotted. Zoom views can
be generated by dragging the mouse over the light curve and clicking on the ``Zoom"
button. The data can be visualized in tabular form by clicking the ``Show Data" button. Finally, the appearance of the plot is customizable (Fig.~\ref{plot}).\\

\end{itemize}

\subsection{VO Discovery tool} To constrain or determine the astrophysical nature of a given objects observed by CoRoT it is necessary, in some cases, to complement the light curve information with other sources of data (physical parameters, spectra, information on the stellar environment, etc). Gathering all this information is a very time-consuming task, especially if performed by hand. 

All the relevant and necessary information already available in astronomical archives and catalogues, however, and can be efficiently retrieved by taking advantage of the Virtual Observatory project. The International Virtual Observatory Alliance (IVOA\footnote{http://ivoa.net}) is an international, community-based initiative to provide seamless access to data available from astronomical archives and services, as well as
to develop state-of-the-art tools for the efficient analysis of this huge amount of information. The Spanish Virtual Observatory (SVO) became part of IVOA in June 2004. LAEX hosts the largest astronomical data centre managed by a Spanish institution (INTA-CSIC) and it is the core of the Spanish Virtual Observatory.

\begin{figure*}[h]
  \centering
\includegraphics[width=12cm,clip=true]{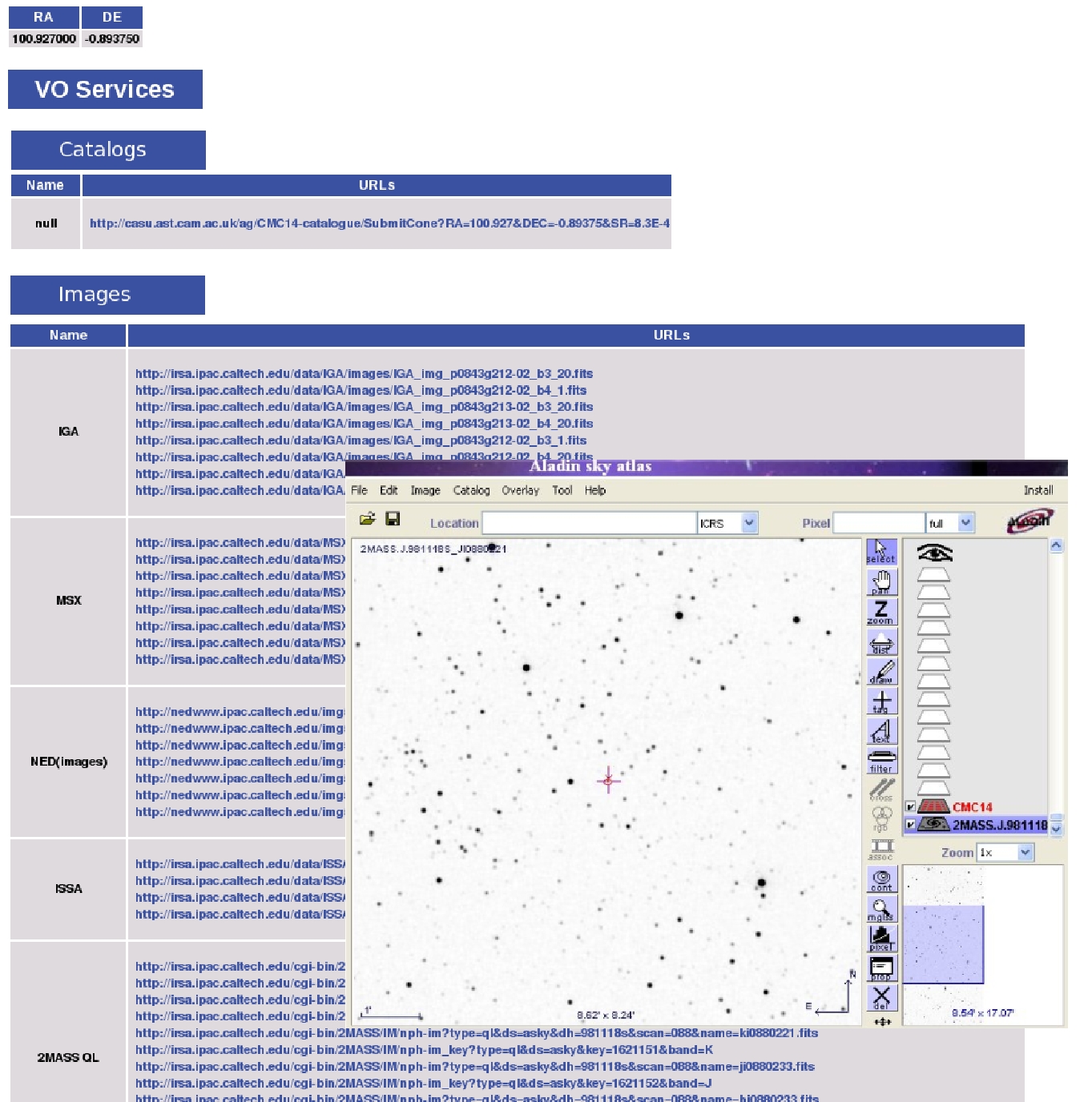}
  \caption{Final list of VO services providing information on the selected target. Images and catalogues can be visualized using Aladin.}
             \label{VOdisc}%
   \end{figure*}

For this purpose, a discovery tool to obtain complementary astrophysical information on CoRoT targets (catalogues, images and spectra) available from VO services has been developed. After querying the National Virtual Observatory (NVO) and European Space Agency (ESA) registries, the tool makes use of the VO monitoring Service\footnote{http://thor.roe.ac.uk/vomon-full/status.xml}. Information retrieved from the operative services is shown as files in VOTable format (an XML-based format defined for the exchange of tabular
data in the context of the Virtual Observatory). These files can be fetched by the user. Images and catalogues can be visualized using Aladin\footnote{http://aladin.u-strasbg.fr/}. Finally, the discovery tool allows the possibility of making ad-hoc queries to archives such as the ones of the Very Large Telescope (VLT), or of the Gemini or Subaru Telescopes (Fig.~\ref{VOdisc}).

\subsection{The VO Service}

Having a VO-compliant archive linked to data mining tools and perfectly integrated in the IVOA structure constitutes an added value of enormous importance for an astronomical project and the optimum scientific exploitation of their datasets.

The Virtual Observatory is primarily concerned with how data are exposed to the world through standardised requests and responses, rather than how they are internally stored, described or manipulated in the archives. Therefore, data providers must comply with VO access protocols to publish their data and data models (abstract entities defining the metadata associated to a given dataset) to the Virtual Observatory to ensure that the same kind of data are described in the same way.

\begin{figure*}[h]
  \centering
\includegraphics[width=13cm,clip=true]{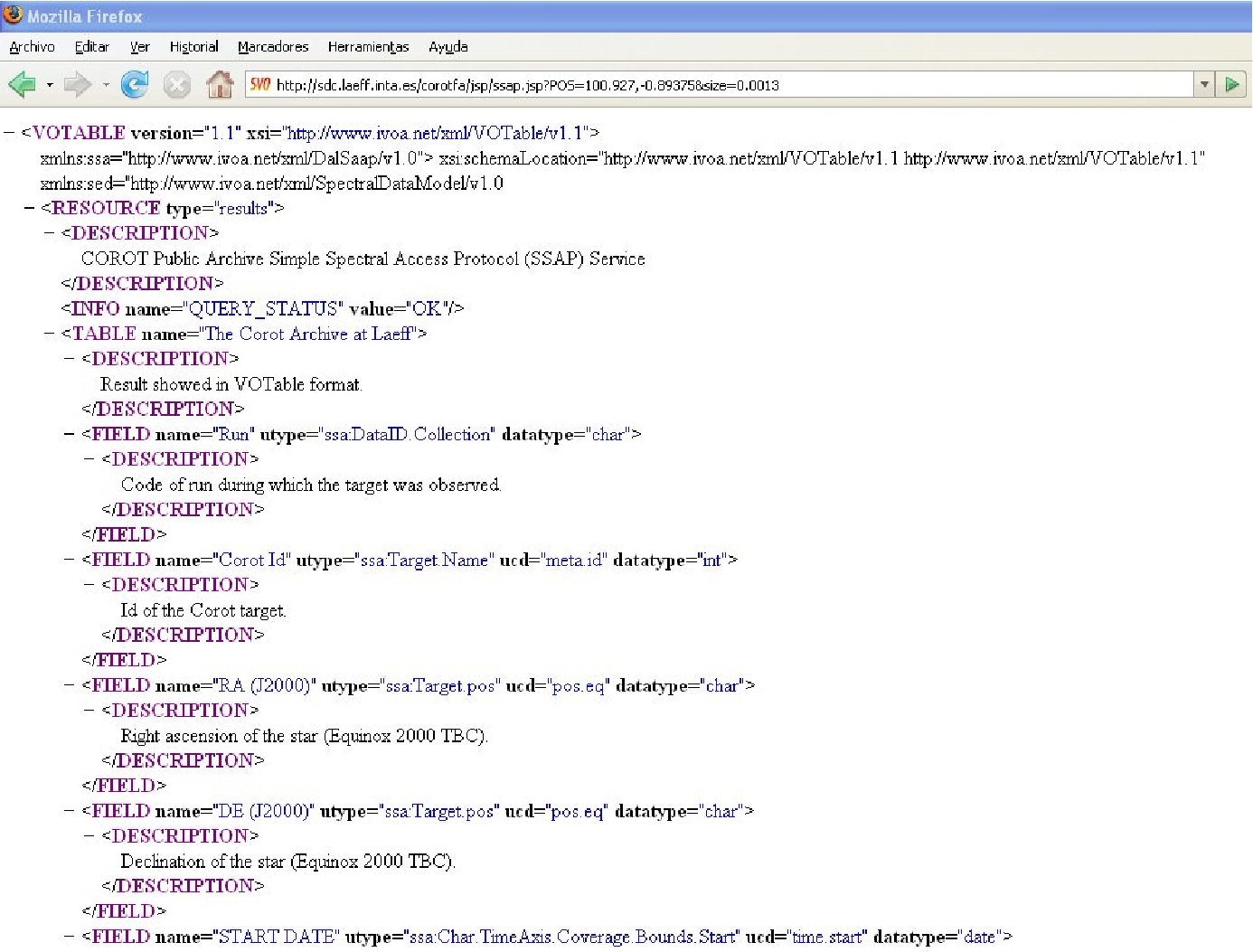}
  \caption{Response of the CoRoT VO service to a SSAP query. The output is in VOTable format.}
             \label{VOserv}%
   \end{figure*}

The CoRoT Public Archive at LAEX was designed following the IVOA standards and requirements. In particular, it implements the SSA (Simple Spectral Access) protocol and its associated data model, a standard
defined for retrieving 1-D data. Through the CoRoT VO service, a client searches for
available data that match certain client-specified criteria using
a HTTP GET request. The response is a table (in VOTable format) describing the available data, including metadata and access references (implemented
as URLs) for retrieving them. (Fig.~\ref{VOserv})

%
\section{Summary}\label{summary}

Archiving constitutes a fundamentally important aspect of any astronomical project that creates data. In this paper we describe two of the CoRoT Public Archives: NStED\footnote{http://nsted.ipac.caltech.edu} and LAEX\footnote{http://sdc.laeff.inta.es/corotfa/}.

The NStED stellar and exoplanet services provide access
to data relevant to exoplanet host stars and bright stars from major
catalogues. The search query tools and cross-referencing capabilities
make NStED a powerful engine for use in connection with exoplanet survey
programs. The NStED Exoplanet transit survey service aims to make time-series photometry data available to the astronomical community in a homogeneous format. The
principal goal is to increase usefulness of the survey data sets by enabling
the extraction of additional scientific results from the data. NStED is continually
updated to reflect the latest results in the literature and to provide
published data access to the broader astronomical community.

NStED will act as the U.S. Portal to the CoRoT data. Its infrastructure is
well matched to both kinds of the CoRoT time-series data products. The
interface is currently being developed to optimize searchability within the
CoRoT data set, provide tools to visualize, manipulate, and characterize
individual light curves, and enable matching of CoRoT data to data contained
in NStED's archive.

CoRoT Public Data will also be available from LAEX as part of the Spanish Virtual Observatory\footnote{http://svo.laeff.inta.es}. The concept of a Virtual Observatory is that all the world's astronomical data should appear as if being stored on the astronomer's desktop, analysable with a user selected workbench of tools and made available through a standard interface. Having a VO-compliant archive represents an added value of enormous importance for an astronomical project as it expands the possibilities of interoperability with other astronomical archives and resources. The development of a discovery tool to gather in an easy and efficient way complementary astrophysical information on CoRoT targets available from VO services is a clear example of the benefits of the VO standardisation.

\begin{acknowledgements}
We would like to express our gratitude to the organizers of the 2009 CoRoT
symposium. The NASA Star and Exoplanet Database is operated by the Jet Propulsion Laboratory, California Institute of Technology, under contract with the National Aeronautics
and Space Administration. NStED wishes to thank the astronomers and research
teams who have generously donated their data sets. This research has made use of the Spanish Virtual Observatory supported
from the Spanish MEC through grants AyA2008-02156, AyA2005-04286 and from the Madrid Regional Government through PRYCIT S-0505/ESP-0361.
\end{acknowledgements}

\end{document}